# An Accurate Current Model for III-V Field Effect Transistors Using a Novel Concept of Effective Transmission Coefficient

*Ehsanur Rahman, Abir Shadman, Sudipta Romen Biswas, Kanak Datta and Quazi D. M. Khosru
ehsaneee@eee.buet.ac.bd

*Abstract*—**In this work, we investigate the transport phenomena in III-V material based buried channel Quantum Well MOSFET with a view to develop a simple and effective model for the device current. Device simulation has been performed in quantum ballistic regime using non equilibrium Green's function (NEGF) formalism. The simulated *I-V* characteristics using a novel concept of effective transmission coefficient has been found to define the reported experimental data with high accuracy. The proposed model has also been effective to capture the transport characteristics reported for other III-V material based field effect transistors. The concept of the proposed effective transmission coefficient and hence the model lends itself to be a simple and powerful device analysis tool which can be extensively used to predict the performance of a wide variety of compound semiconductor devices in the pre-fabrication stage. It has also demonstrated consistency with device characteristics for doping concentration and channel length scaling. Thus the model can help the device/process engineers to tune the devices for the best possible performance.**

## I. INTRODUCTION

The ever expanding demand of ultra-fast state of the art electronic devices requires higher computational power which has necessitated the scaling of MOS transistors into nano meter regime over the last decade – a trend embodied in Moore's law. Continued scaling of planar, silicon, CMOS transistors will be more and more difficult because of both fundamental limitations and practical considerations as the transistor dimensions approach deeply scaled nanometer regime[1]. Degraded short channel effects, gate leakage and parasitic components have all pushed the Si CMOS to their fundamental limit. Motivation for consideration of III-V materials in traditional MOS transistor structures for digital applications is their extremely high electron mobility due to small effective mass in the Γ valley. High electron mobility of III-V materials has made them suitable for low power, high frequency operations. III-V material based Field Effect Transistors are considered as potential candidate to sustain transistor scaling beyond 10 nm node while maintaining improved device performance. To predict the potential benefit of III-V devices over conventional Si MOSFETs and further optimize their operation, deep understanding of device physics is necessary. Modelling and simulation can offer useful insight in this regard. III-V material based buried channel Quantum Well device structure has been reported in recent past [2]. Despite numerous experimental works on hetero-junction quantum well structures, simulation based *I-V* characterization and benchmarking with experiment is rarely available in the literature. The most crucial part of any simulation-based study involves explaining experimental data using device physics and carrier transport phenomena.

The possible physics behind carrier transport in a device can be extracted by explaining the experimental current using an effective computationally optimized simulator. In this work, a highly efficient, two-dimensional quantum ballistic simulator has been used. NEGF formalism based on effective masses in mode space has been utilized [3], [4] for quantum well structures which is evolved from nanoMOS2.5 [5]. Using the same approach [5], quantum ballistic simulation has been performed for the structure demonstrated by Lin et. al. [2]. Here a novel concept of effective transmission coefficient has been introduced for explaining simulation current with experiment. This new technique has resulted in excellent agreement with reported experimental data for different structures of III-V material based Field Effect Transistors.

## II. EFFECTIVE TRANSMISSION COEFFICIENT

In practice, device current is calculated by integrating the multiplication of Fermi function difference between source and drain and transmission coefficient (both function of energy) over a certain range of energy. NEGF basically produces step like transmission coefficient resulting in a device current that exhibits high disagreement with the measured current. Here, instead of step like transmission coefficient, we introduce an effective transmission coefficient in the form of Gaussian-like distribution over an appropriate range of energy. It is the very essence of the model that it approximates various non ideal effects in a real device in a very simplified way without any rigorous and computationally expensive analysis while maintaining significant accuracy. In this work, our model integrates various real transport phenomena in a simplified way through Gaussian distribution coefficient. An empirical expression for the Gaussian like energy distributed effective transmission coefficient is proposed as:

$$T(E) = K\, exp^{\frac{-\left(E-E_{cp}-E_{of}\right)^2}{\left(2*\left((a+b*V_d)\right)^2\right)}}$$

Here $K$ is the maximum transmission coefficient which has been kept 1 in this work, $T\ (E)$ is energy resolved transmission coefficient, $E$ is carrier energy, $E_{cp}$ is top of the energy barrier from source to drain which is obtained by solving Schrodinger equations along the confinement direction, $E_{of}$ is Gaussian peak shifter and $V_d$ is drain voltage. $a$ and $b$ are fitting parameters where $a$ is unit less, $b$ is in $V^{-1}$. In the above equation '$b$' and '$E_{of}$' are device dependent free fitting parameters, while the parameter

'*a*' shows approximate linear dependence on gate voltage for explaining experimental data with good agreement. From the empirical relation it is found that transmission about top of the barrier dominates the device current. Contribution from electron flux at other energy levels drops off drastically giving a transmission probability density function with very small variance. The denominator of the Gaussian determines the broadening of the distribution. The term $(a+b*V_d)$ increases with $V_d$ because with increase of $V_d$, more and more higher energy electrons take part in conduction. This concept is unique in the sense that the experimental current can be explained simply by tuning the transmission coefficient of electron at different energy levels around the top of the barrier. After obtaining the effective transmission coefficient the final current is determined by the following equation

$$I = \frac{2q}{h}\int T(E) * (Fs - Fd)dE$$

Where $q$ is electron charge, $h$ is plank's constant, $Fs$ and $Fd$ are source and drain Fermi functions respectively.

## III. Device Structure and Simulation

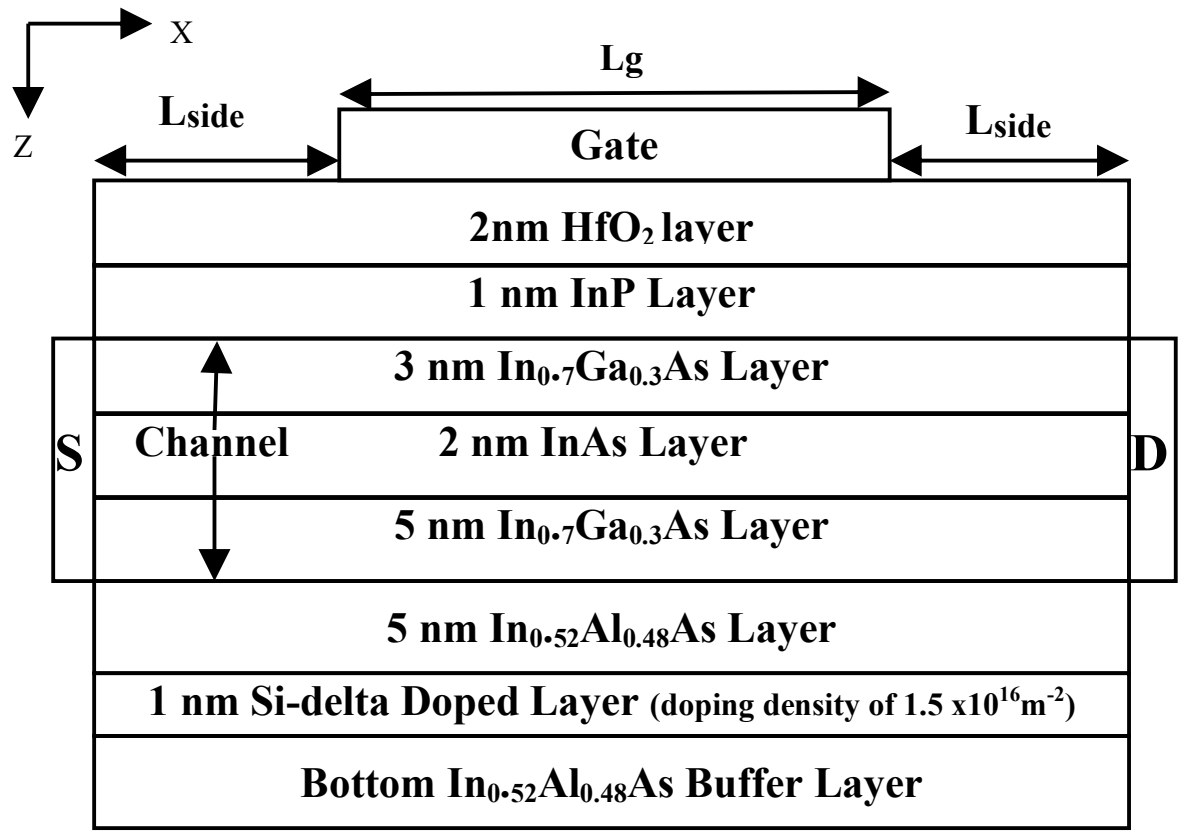

Fig. 1. Schematic view of the simulation region of the Quantum Well device used in this study. Here Lg=30 nm and Lside is kept at 44 nm.

Fig.1 illustrates the device structure and dimensions used in the simulation. It has a $In_{0.7}Ga_{0.3}As/InAs/In_{0.7}Ga_{0.3}As$ composite channel as demonstrated by Lin et. al. [2]. In real devices, two-dimensional current flows through the doped heterostructure stack from the raised source to the drain. Ideal contacts have been placed on two sides instead of real contact and its associated resistance for the simplicity of computation. Under feasible gate and drain bias, charge has been calculated using NEGF method [7] based on effective masses in mode space. Self-consistent simulation is done using coupled Schrödinger-Poisson equation [8]. Energy resolved transmission coefficient has been tuned to a Gaussian distribution in place of step like function (as shown in Fig. 2) as a post processing after achieving convergence in coupled Schrodinger Poisson solution.

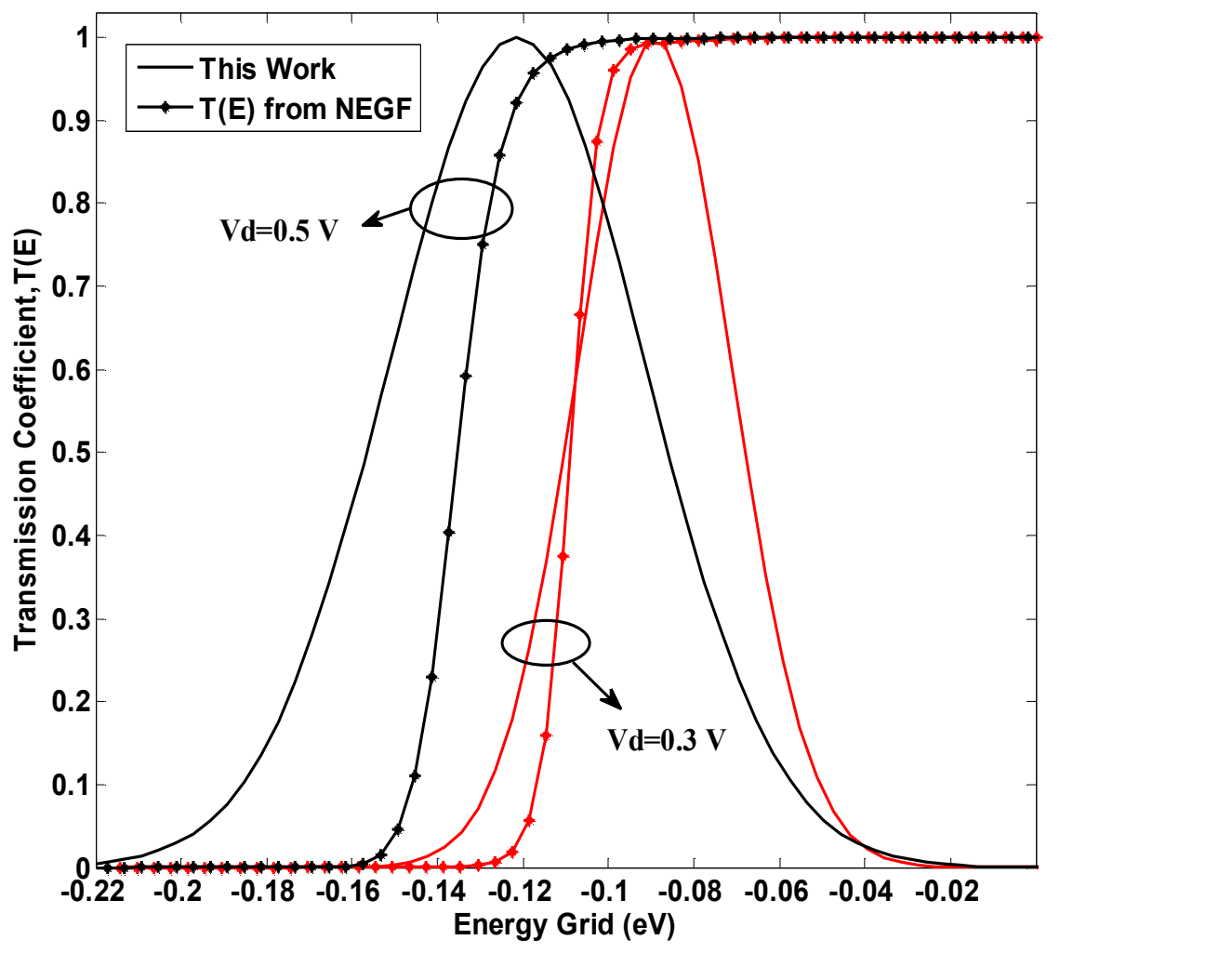

Fig. 2. Comparative illustration of the proposed effective transmission coefficient with the step like transmission coefficient from NEGF. Results are shown for Vg=0.3V.

## IV. Interpretation of the Model

Transmission coefficient indicates the probability of an electron at a particular energy to transmit through potential barrier from source to drain. In this simulation, transmission coefficient has been modeled to a Gaussian shape to incorporate deemed non-ideal effects which are inherent in experimental data. Results thus obtained are shown in Fig. 2 for the device under investigation. From NEGF, we get step like transmission probability as shown in Fig. 2. But in reality, many phenomena such as scattering and quantum mechanical tunneling through the barrier will broaden the step-like transition near top of the barrier. So, electron flux below the top of barrier should contribute to current as can be seen in Fig. 3 which presents a NEGF simulation incorporating tunneling.

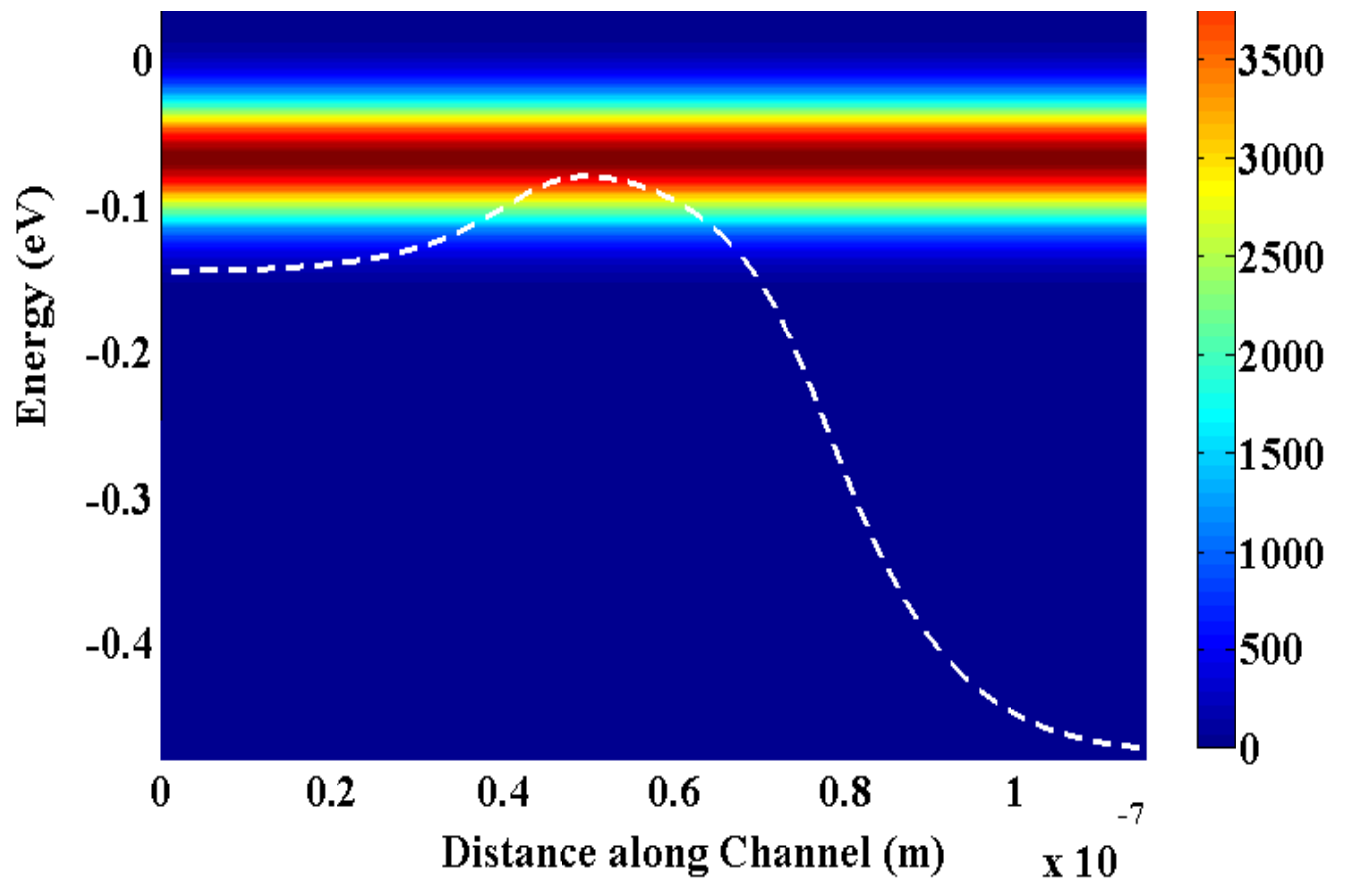

Fig.3: Simulated energy resolved current (A/m) from NEGF for the device structure shown in Fig.1 at Vg=0.3 V and Vd=0.3 V. The color scale has a unit of A/m.

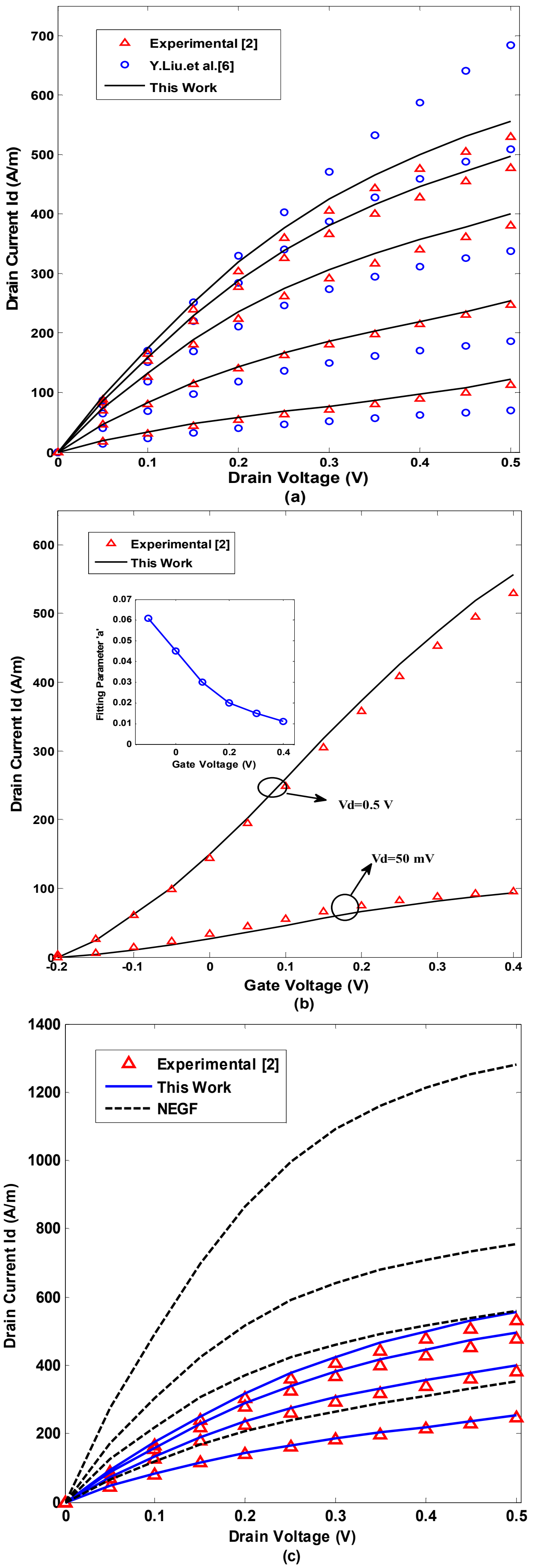

Fig. 4: Transport characteristics as obtained using the proposed model. (a) Comparison of *Id-Vd* characteristics of the proposed method with experimental data [2] and Liu.et. al [6] with Gate Voltage varying from 0 to 0.4V in step of 0.1V. (b) Comparison of *Id-Vg* characteristics with the experimental data [2]. In producing both the transport characteristics we use the proposed effective transmission coefficient keeping '$b$'(= 0.03 v$^{-1}$); '$E_{of}$'(= 0.02 eV) constant and varying '$a$' with gate voltage as shown in the inset of Fig. b. (c) Comparison of transport characteristics obtained from NEGF and this work with experimental data[2] for the same device with Gate Voltage varying from 0.1 to 0.4V in step of 0.1V

Physically, current contribution from electron at very high energy above top of the barrier is also negligible which is evident from Fig. 3. The peak of the transmission coefficient is shifted towards lower energy with increasing drain bias which is consistent with the result presented in Fig. 2 as well as the well-known short channel effect called drain induced barrier lowering.

## V. Comparison Between Effctive Transmission Coefficient and Contact Resistance Tuning

Using the proposed concept of effective transmission coefficient current-voltage (*I-V*) characteristics are simulated for the device illustrated in Fig.1. *Id-Vd* characteristics are also produced including series resistances at the source and drain ends in the quantum ballistic simulation as reported by Liu et al. [6]. Figure 4(a) exhibits a comparison between the simulation results with respect to experimental *I-V* characteristics for the same device. It demonstrates a good agreement between the proposed model and the experimental results. *Id-Vg* characteristics simulated with the proposed model also exhibit good agreement with the experimental data as presented in Fig. 4(b). For the same device, Fig. 4(c) compares the *Id-Vd* characteristics obtained from NEGF and from the proposed model with experimental data. It is evident from Fig. 4(c) that current obtained from NEGF is almost 2 to 3 times higher compared to experimental current in high gate bias regime. Hence, transport characteristics based solely on NEGF gives a very poor prediction of device performance. The proposed model has been consistent in explaining the transport phenomena in HEMT structure [9] and other III-V surface channel quantum well field effect device [10] as demonstrated in Fig. 5. It can be noticed in Fig. 5(a) that the contact resistance technique [6] defines the transport characteristics in HEMT structure with good accuracy for low gate voltage (~0.3V) regime. However, appreciable disagreement with the experimental data becomes apparent beyond the low voltage regime. This may be due to the fact that it is difficult to explain experimental current using simulation for all gate voltages by using a simple contact resistance without incorporating other physics like scattering and tunneling which govern carrier transport. From the results presented in Fig. 4 and Fig. 5 it is worth mentioning that the proposed concept of effective transmission coefficient is an efficient tool for explaining transport characteristics in III-V material based nanoscale devices for the practical range of gate and drain voltages.

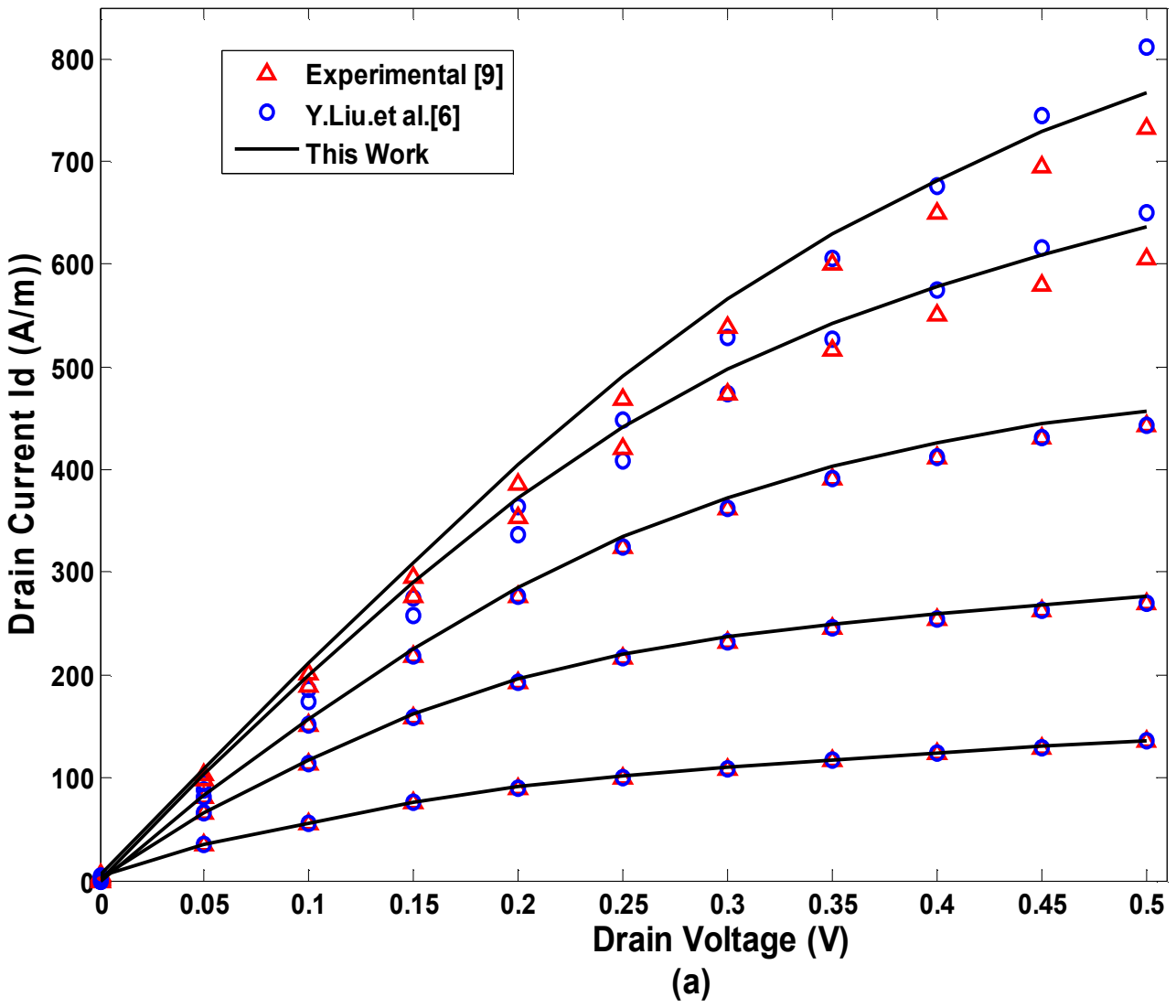

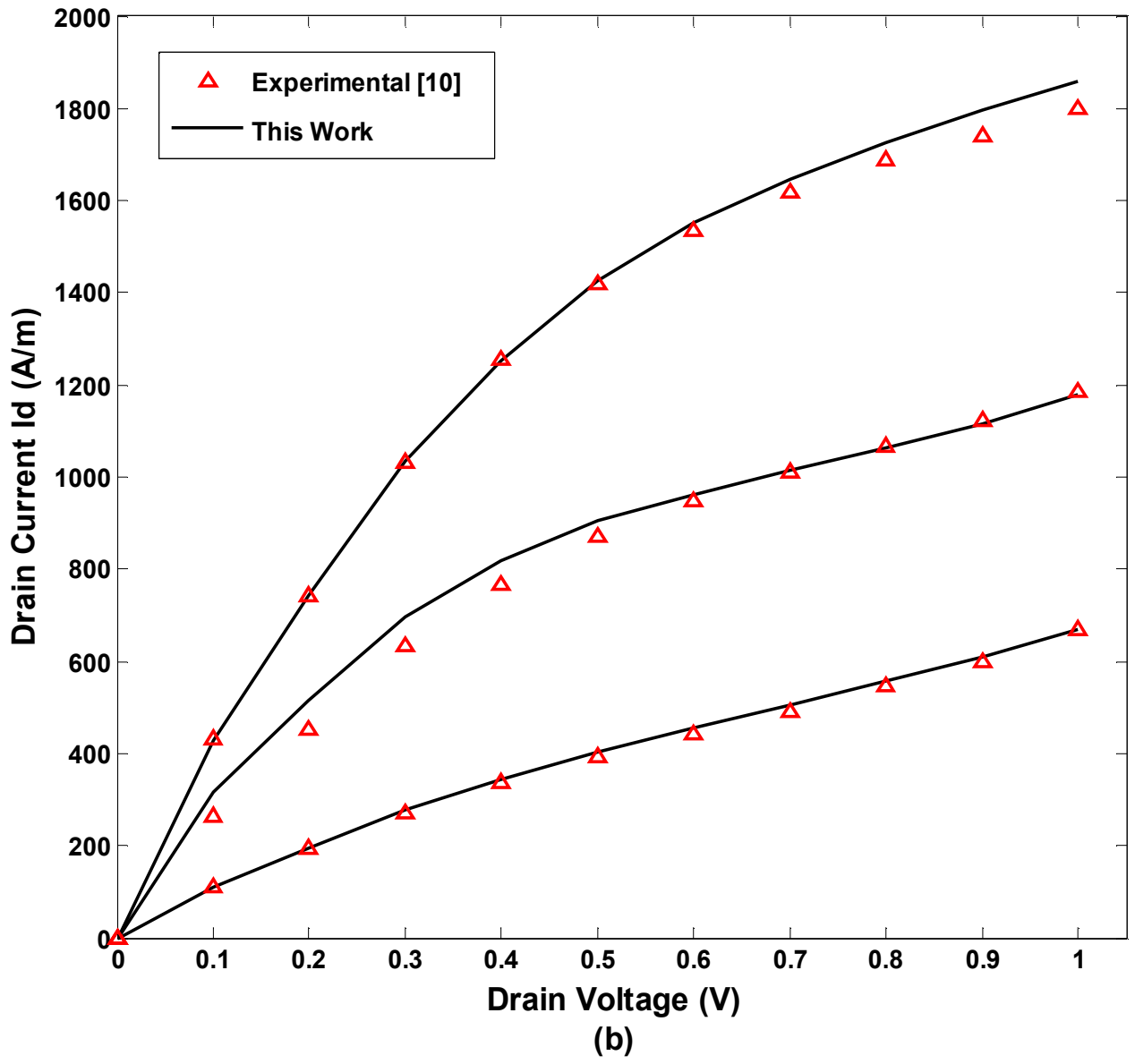

Fig.5: (a) Comparative illustration of *Id-Vd* characteristic for the III-V HEMT structure reported by Liu.et al. [9] for $L_g$=40 nm with Gate Voltage varying from 0.1 to 0.5V in step of 0.1V. (b) Comparative illustration of *Id-Vd* characteristic for the III-V surface channel Quantum Well structure reported by Egard.et al. [10] for $L_g$=55nm with Gate Voltage varying from 0.6 to 1.4V in step of 0.4V.

## VI. Performance Dependence on Device Channel length and Doping Density

Another significant feature of the proposed model is that it can qualitatively predict device performance for various device parameters like channel length and doping density. Fig. 6 shows the drain current for various doping densities and channel lengths for the device structure illustrated in Fig. 1 using the same coefficient values reported in Fig.4. The results presented in Fig. 6 demonstrate that the model predicts device performance for various parametric variations. However, even though the experimental data for such variations has not been reported yet and thus it is not possible to benchmark the simulator for those variations, it shows that the simulator is flexible enough to be used for various parametric variations and it follows the general trend consistent with the device physics. Thus, it can be a useful tool to optimize the device performance from design point of view.

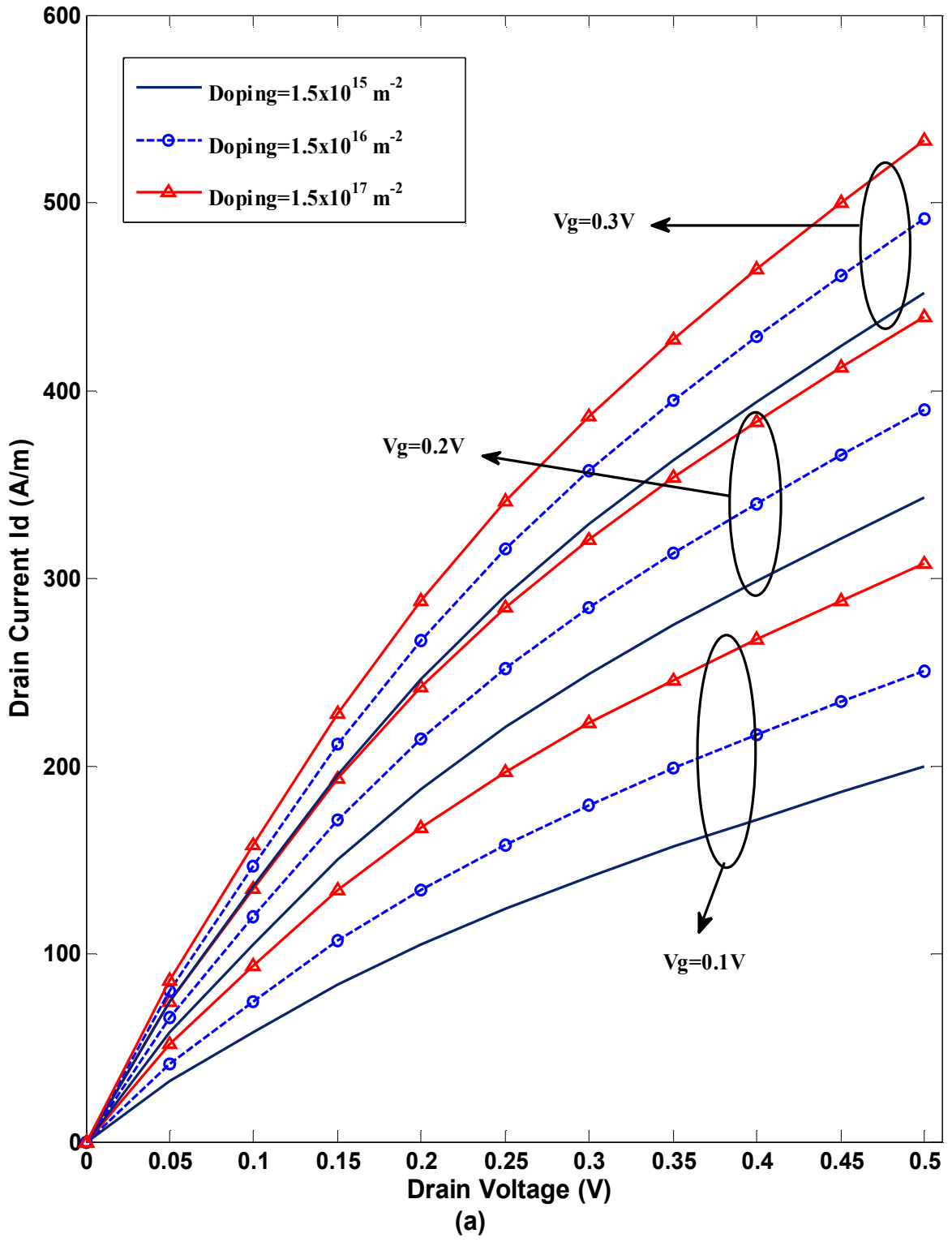

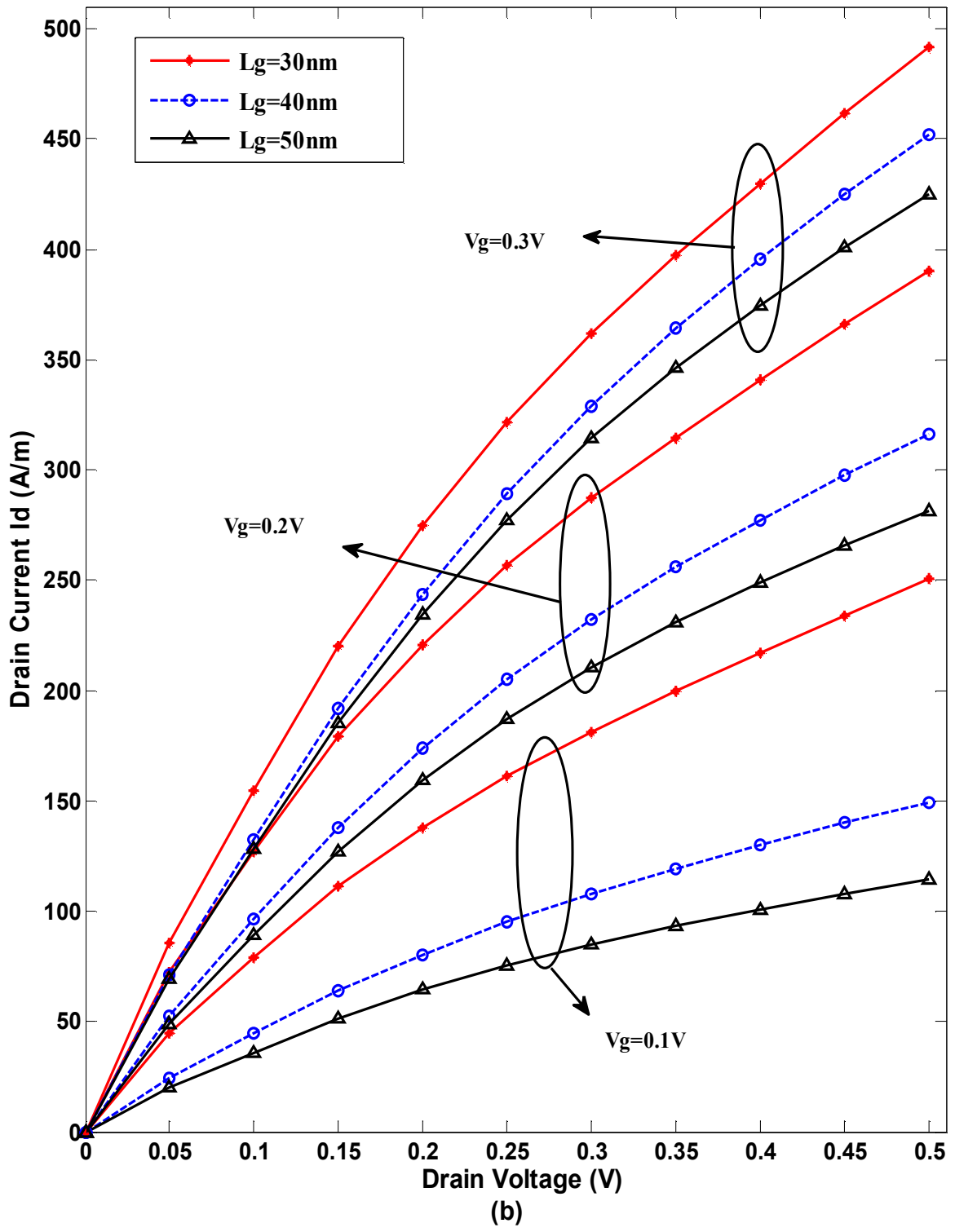

Fig. 6: (a) *Id-Vd* characteristics for various doping densities of the device illustrated in Fig.1. with Lg=30 nm (b) *Id-Vd* characteristics of the same device for various channel lengths with doping density of 1.5 $x10^{16}$ $m^{-2}$

## VII. Conclusion

A novel concept of effective transmission coefficient has been introduced to explain the transport characteristics in III-V material based field effect transistors of various structures. NEGF formalism based carrier transport mechanism in quantum ballistic regime has been utilized in core simulation structure. The use of uncoupled mode space approach has greatly reduced the computational burden compared to real space approach. The transmission coefficient of electrons from source to drain has been found to exhibit Gaussian like distribution which indicates that the energy resolved current density inside the device follows normal distribution. The proposed model has been successfully used to predict device performance for various parametric variations like doping density and channel length. The model presented in this work is a unique device analysis tool that is simple and effective to design state of the art III-V material based field effect transistors for optimum performance.